\documentclass[aps,prl,twocolumn,amsmath,amssymb,floatfix]{revtex4-1}
\usepackage{amsmath}
\usepackage{amssymb}
\usepackage{amsfonts}
\usepackage{listings}
\usepackage{enumerate}
\usepackage{latexsym}
\usepackage{multirow}
\usepackage{graphicx}
\usepackage{braket}
\usepackage[pdftex,colorlinks=false]{hyperref}
\usepackage{xcolor, colortbl}
\usepackage{verbatim}

\usepackage{times}
\usepackage{array, makecell}
\usepackage[export]{adjustbox}
\usepackage{tabularx}

\newcolumntype{M}[1]{>{\centering\arraybackslash}m{#1}} 

\newcolumntype{N}{@{}m{0pt}@{}}

\newcommand{\beq}{\begin{equation}}
\newcommand{\eneq}{\end{equation}}
\newcommand{\bs}[1]{\boldsymbol{#1}}

\def\be{\begin{equation}}
\def\ee{\end{equation}}
\def\ba{\begin{eqnarray}}
\def\ea{\end{eqnarray}}

\def\R{{\rm Re}}
\def\Z{\mathbb{Z}}
\def\C{\mathbb{C}}

\def\dag{\dagger}

\renewcommand{\vec}{\bs}
\newcommand{\kk}{\vec{k}}

\def\beq{\begin{equation}}
\def\eeq{\end{equation}}
\def\barray{\begin{eqnarray}}
\def\earray{\end{eqnarray}}

\font\upright=cmu10 scaled\magstep1
\def\stroke{\vrule height8pt width0.4pt depth-0.1pt}

\def\Zmath{\mathbb{Z}}
\def\Qmath{\vcenter{\hbox{\upright\rlap{\rlap{Q}\kern
                   3.8pt\stroke}\phantom{Q}}}}
\def\Nmath{\vcenter{\hbox{\upright\rlap{I}\kern 1.7pt N}}}
\def\Cmath{\vcenter{\hbox{\upright\rlap{\rlap{C}\kern
                   3.8pt\stroke}\phantom{C}}}}
\def\Rmath{\vcenter{\hbox{\upright\rlap{I}\kern 1.7pt R}}}
\def\Z{\ifmmode\Zmath\else$\Zmath$\fi}
\def\Q{\ifmmode\Qmath\else$\Qmath$\fi}
\def\N{\ifmmode\Nmath\else$\Nmath$\fi}
\def\C{\ifmmode\Cmath\else$\Cmath$\fi}
\def\R{\ifmmode\Rmath\else$\Rmath$\fi}

\input{epsf}

\newcounter{defcounter}
\begin{document}

%
%
%

\title{Atomic limit and inversion-symmetry indicators for topological superconductors}

\author{
Anastasiia~Skurativska}
\affiliation{
 Department of Physics, University of Zurich, Winterthurerstrasse 190, 8057 Zurich, Switzerland
}
\author{
Titus~Neupert}
\affiliation{
 Department of Physics, University of Zurich, Winterthurerstrasse 190, 8057 Zurich, Switzerland
}
\author{
Mark~H~Fischer}
\affiliation{
 Department of Physics, University of Zurich, Winterthurerstrasse 190, 8057 Zurich, Switzerland
}

\begin{abstract}
	Symmetry indicators have proven to be extremely helpful in identifying topologically non-trivial crystalline insulators using symmetry-group representations of their Bloch states. 
	An extension of this approach to superconducting systems requires defining an appropriate atomic limit for Bogoliubov-de-Gennes Hamiltonians.  Here, we introduce such a notion of atomic limit and derive  a $\mathbb{Z}_{2^d}$-valued symmetry indicator for inversion-symmetric superconductors of $d$ dimensions.
	This indicator allows for a refined topological classification including higher-order phases for systems in the superconducting symmetry classes D and DIII.
	We further elucidate the bulk-boundary correspondence of these phases using Dirac surface theories.
	Requiring only the normal-state band structure and the superconducting order-parameter symmetry as an input, this indicator is well suited for a search of topological superconductors using first-principles calculations.
\end{abstract}

\date{\today}

\maketitle

\textit{Introduction} --	 
Unconventional, and in particular topological superconductors (TSC) are of fundamental research interest, due to the existence of gapless Majorana modes on their boundaries \cite{kitaev:2001}. 
These Majorana modes are anyons that obey non-Abelian statistics and are topologically protected, meaning that they cannot be removed from the boundary without breaking their protecting symmetries or closing the gap \cite{kitaev:2001,readgreen:2000,ivanov:2001}. Unlike topological insulators, however, only few candidate TSCs exist \cite{sato:2017}.

Conveniently, Majorana modes are the only anyons that admit a description in terms of free fermions. Thus, a description of TSCs in terms of quadratic Bogoliubov-de-Gennes (BdG) Hamiltonians suffices for most purposes. These Hamiltonians always posses particle-hole symmetry (PHS) by construction. In the following, we are interested in topological phases of generic spin-orbit coupled TSCs with or without additional time-reversal symmetry (TRS), which correspond to symmetry classes D and DIII in the Altland Zirnbauer (AZ) scheme~\cite{schnyder:2008, ryu:2010}, respectively. 
Absent any further symmetries, both classes support gapped topological superconducting phases in one dimension (1D) and two dimensions (2D), while in three dimensions (3D), only class DIII allows for such a phase.
This topological classification can be refined by considering addition non-local (spatial) symmetries like inversion, rotations, or reflection. In the field of topological insulators, this refinement led to the discovery of topological crystalline insulators \cite{fu:2011,hsieh:2012,ando:2015,cornfeld:2019}, and likewise superconductors with gapless modes protected by spatial symmetries, topological crystalline superconductors \cite{ando:2015,cornfeld:2019,shiozaki:2014,liu:2019}. 

Two recent developments motivate our study: First, the bulk-boundary correspondence for  topological crystalline insulators and superconductors is very rich, including corner states in 2D and hinge as well as corner states in 3D, instead of edge and surface states, respectively. Systems with such a generalized bulk-boundary correspondence are termed higher-order topological insulators/superconductors \cite{benalcazar:2017,schindler:2018,geier:2018,trifunovic:2019,khalaf:2018a,khalaf:2018b,lee:2019,scheng:2019,langbehn:2017,wang:2018,yanase:2019}.
Second, several advances have been made to simplify the determination of topological phases from bulk electronic properties using group representation approaches~\cite{po:2017,bradlyn:2017,khalaf:2018b}.
In particular, symmetry indicators based on early work of Fu and Kane~\cite{fu:2007} allow to distinguish topological phases through the transformation properties of the occupied electronic states, rather than lengthy calculations of topological indices. These properties are easily accessible from first-principles calculations using density functional theory (DFT).

The basic idea of symmetry indicators is to quantify the possible mismatch between the real space representation in terms of localized Wannier orbitals and the momentum representation of (Bloch) bands~\cite{po:2017, khalaf:2018b}. Importantly, the atomic insulator, in other words localized Wannier orbitals without mutual overlap, defines an atomic limit (AL) of trivial Bloch bands.
Note that the proper definition of an atomic limit for insulators allows to promote the inversion-symmetry indicator to a $\mathbb{Z}_4$ quantity in three dimensions~\cite{khalaf:2018b}, compared to the $\mathbb{Z}_2$ valued Fu-Kane indicator~\cite{fu:2007}. 
Group representation approaches have been successful in predicting topological crystalline insulators and higher-order topological insulators from material databases~\cite{fengtang:2019,materiae:2019,vergniory:2019}.

In order to adapt an analogous approach of symmetry indicators for TSCs that can be applied within a DFT approach, one faces several challenges. First, the band structure of a material in the superconducting phase is not available from DFT calculations. A useful symmetry indicator, thus, should be defined in terms of the symmetry representations of the normal-metal bands in addition to the order-parameter symmetry, rather than those of the BdG Hamiltonian. Further, on a practical level, the fact that the electronic spectrum is generically not bounded from above means that the particle-hole-doubled BdG band structure is not bounded from below. These problems have indeed been addressed in previous work~\cite{ono:2018tmp}.
Finally, the superconducting spectrum of the BdG Hamiltonian is the quasi-particle excitation spectrum and not the electron excitation spectrum as in the case of insulators. Therefore, the AL for superconductors needs to be specified. 

In this work, we show that the irreducible representations of the BdG band structure alone are insufficient to formulate a comprehensive symmetry indicator for inversion-protected topological superconductors. 
Instead, we introduce an atomic limit for superconductors in terms of the normal metal bands and with respect to that we define the symmetry indicators for higher-order inversion-symmetric TSC in classes D and DIII. 
Unlike the symmetry indicators for topological insulators, these indicators are $\mathbb{Z}_8$ valued in 3D. We show that the symmetry indicators are consistent with the existing topological invariants for the strong TSC phases and agree with the previous results regarding the Fermi surface topology~\cite{sato:2009a, sato:2010, fu:2010, qi:2010}. Additionally, they hold the information about higher order TSC.
We demonstrate the predicted higher order topological bulk-boundary correspondence with surface Dirac Hamiltonians. 

 \begin{table}
 	\centering
		\begin{tabular}{M{0.3cm}|m{0.5cm}!{\vrule width 1.25pt}cc|cc|cc} 
			 & & \multicolumn{2}{c|}{$n = 1$} & \multicolumn{2}{c|}{$n = 2$} & \multicolumn{2}{c}{$n = 3$}\\
			    $d$ & $\kappa_{d, \xi}$ &DIII  &  D & DIII & D & DIII &D\\
			\noalign{\hrule height 1.25pt}
			 \vspace{4pt}
			3 & $\mathbb{Z}_8$ &  \includegraphics[width=1cm, valign=c]{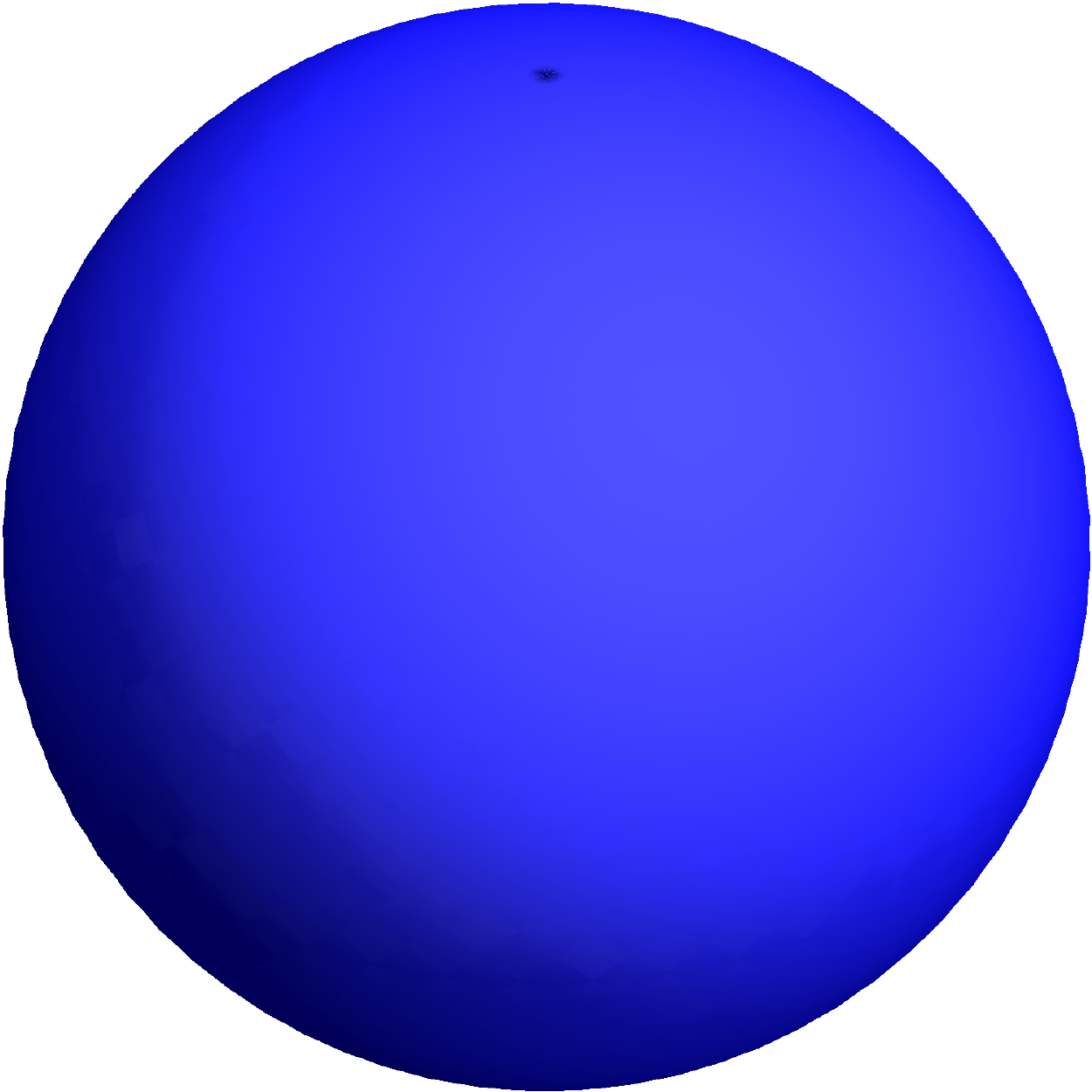} &\includegraphics[width=1cm, valign=c]{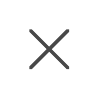} &  \includegraphics[width=1cm, valign=c]{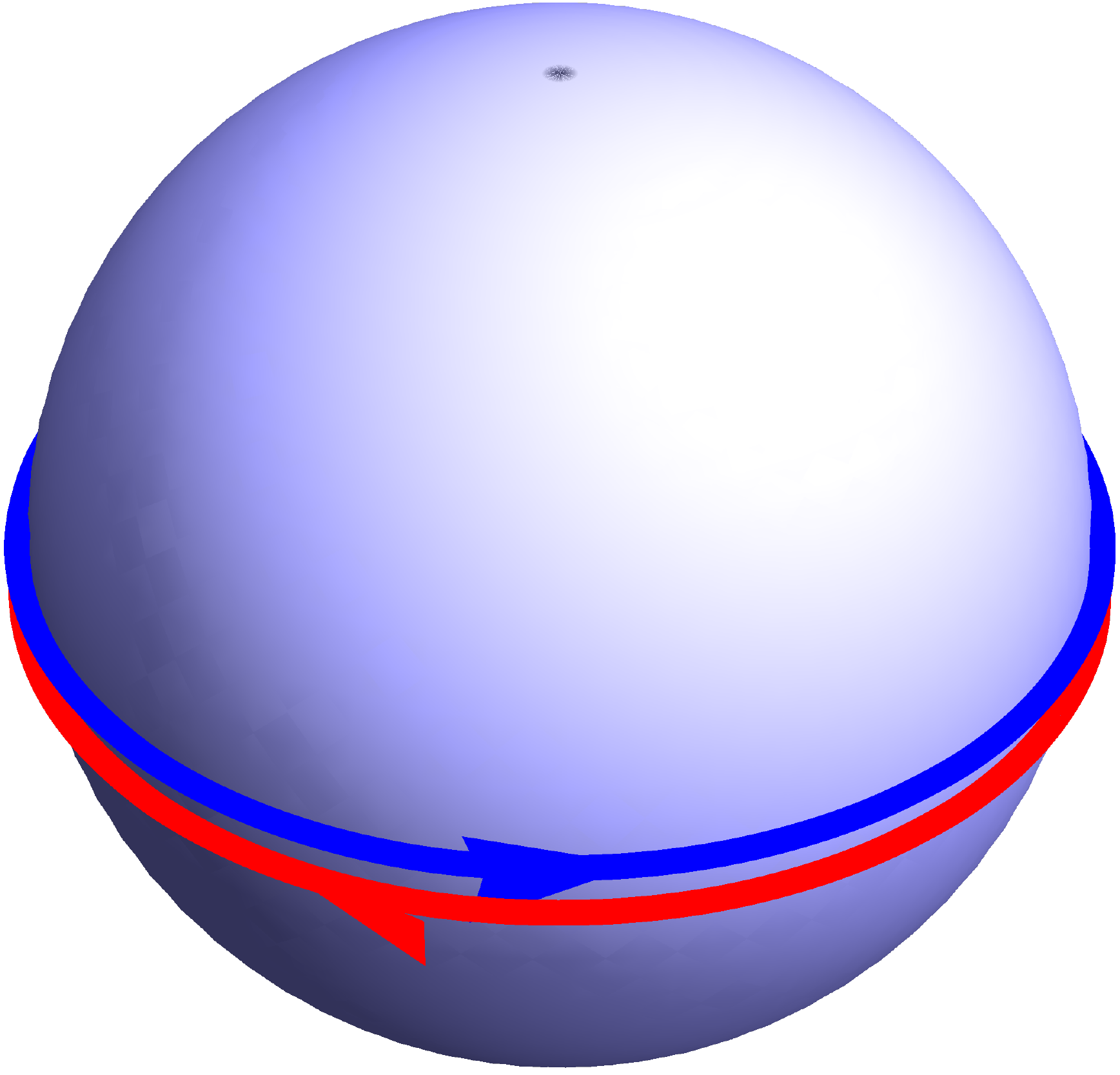}  & \includegraphics[width=1cm, valign=c]{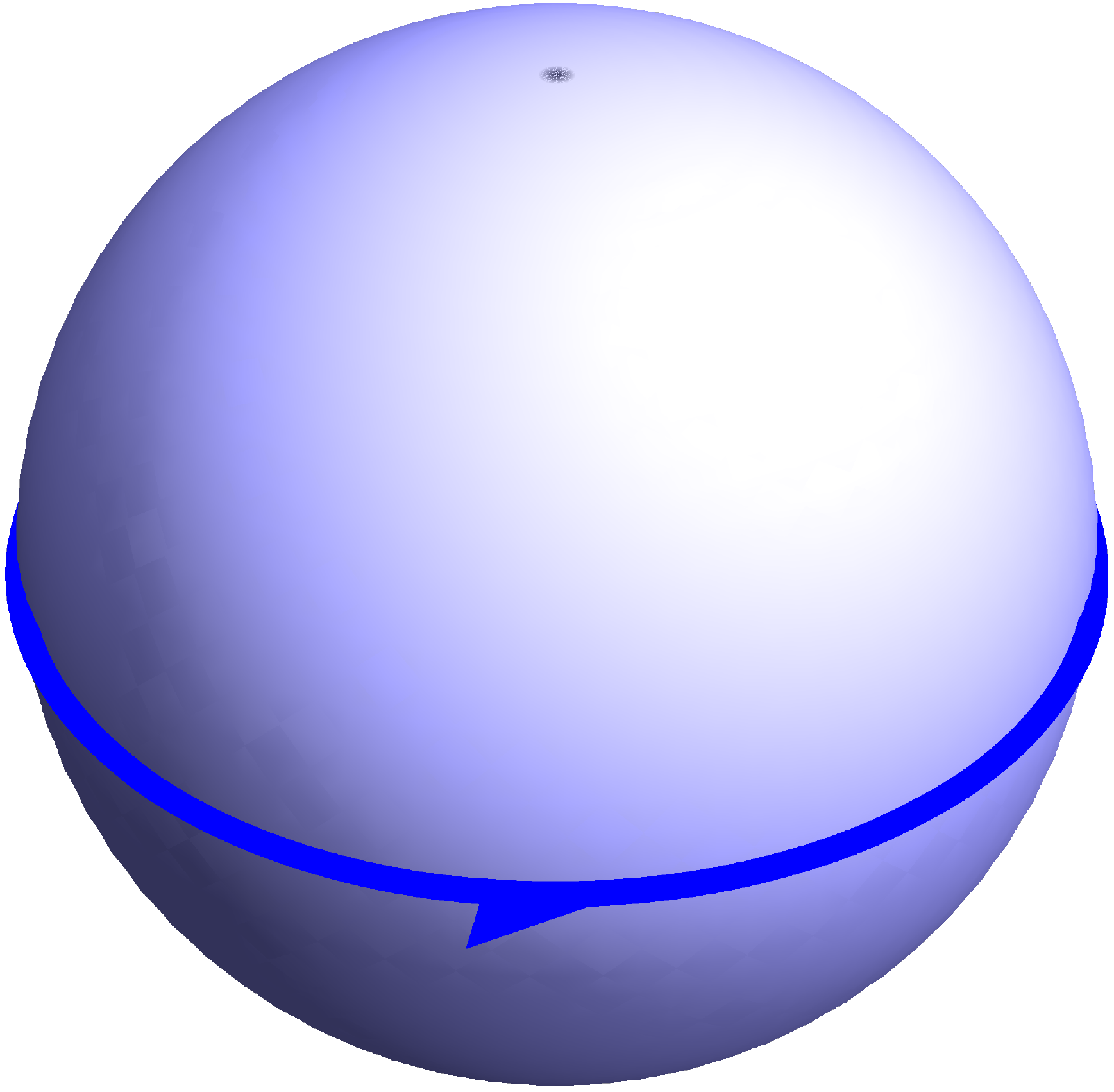} &  \includegraphics[width=1cm, valign=c]{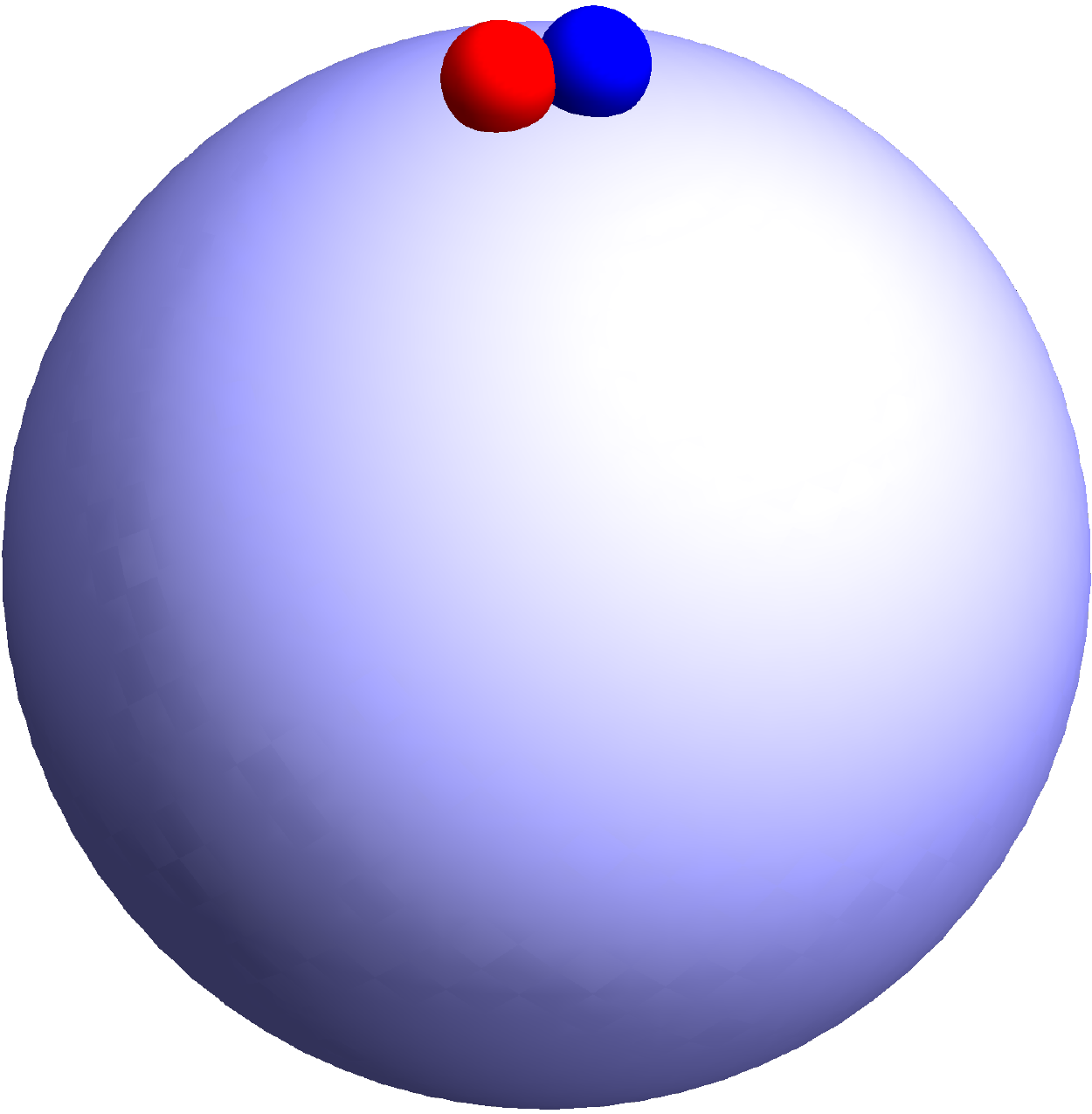}   &\includegraphics[width=1cm, valign=c]{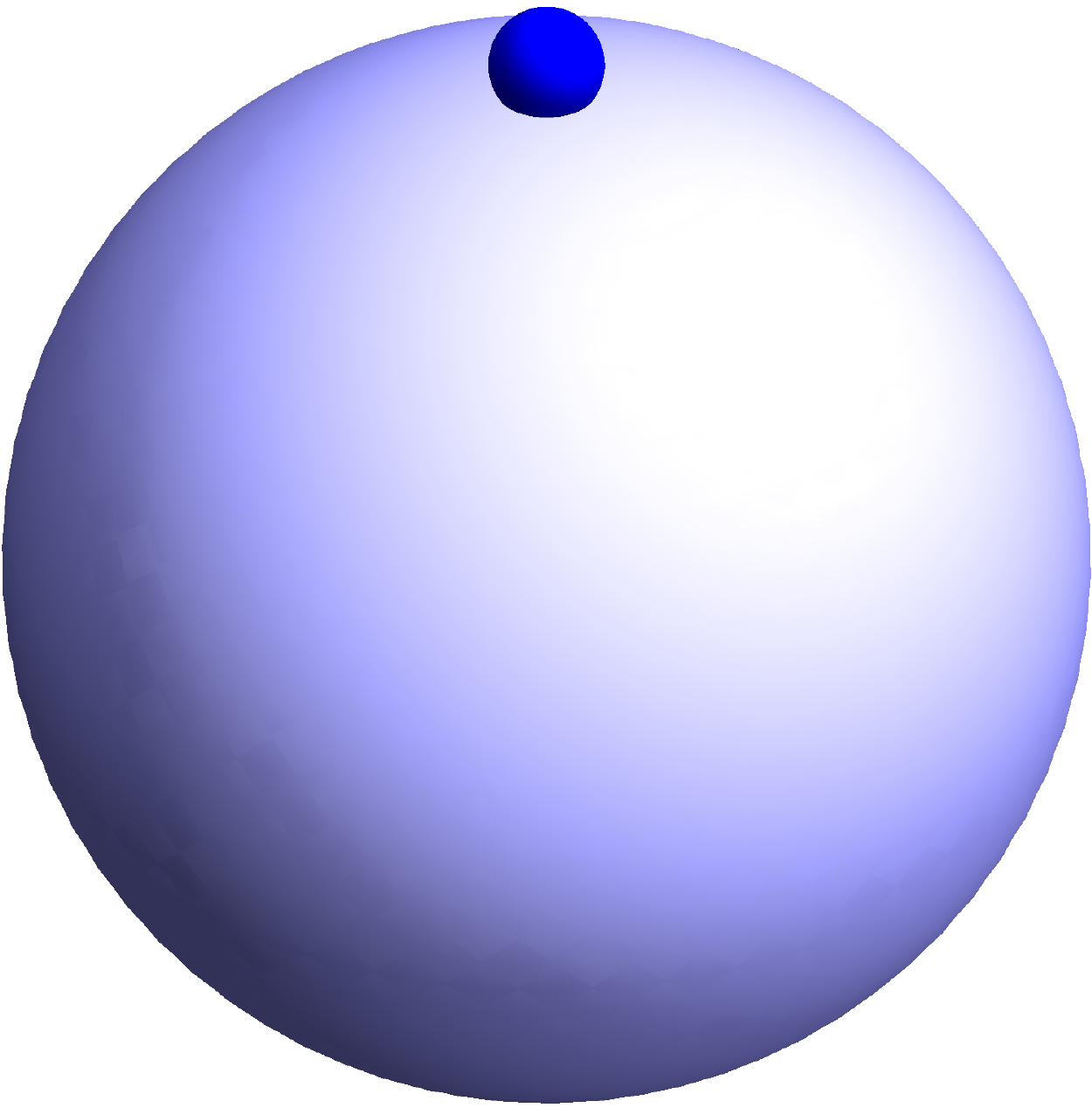}\\
			\hline
			2 & $\mathbb{Z}_4$ &  \includegraphics[width=1cm, valign=c]{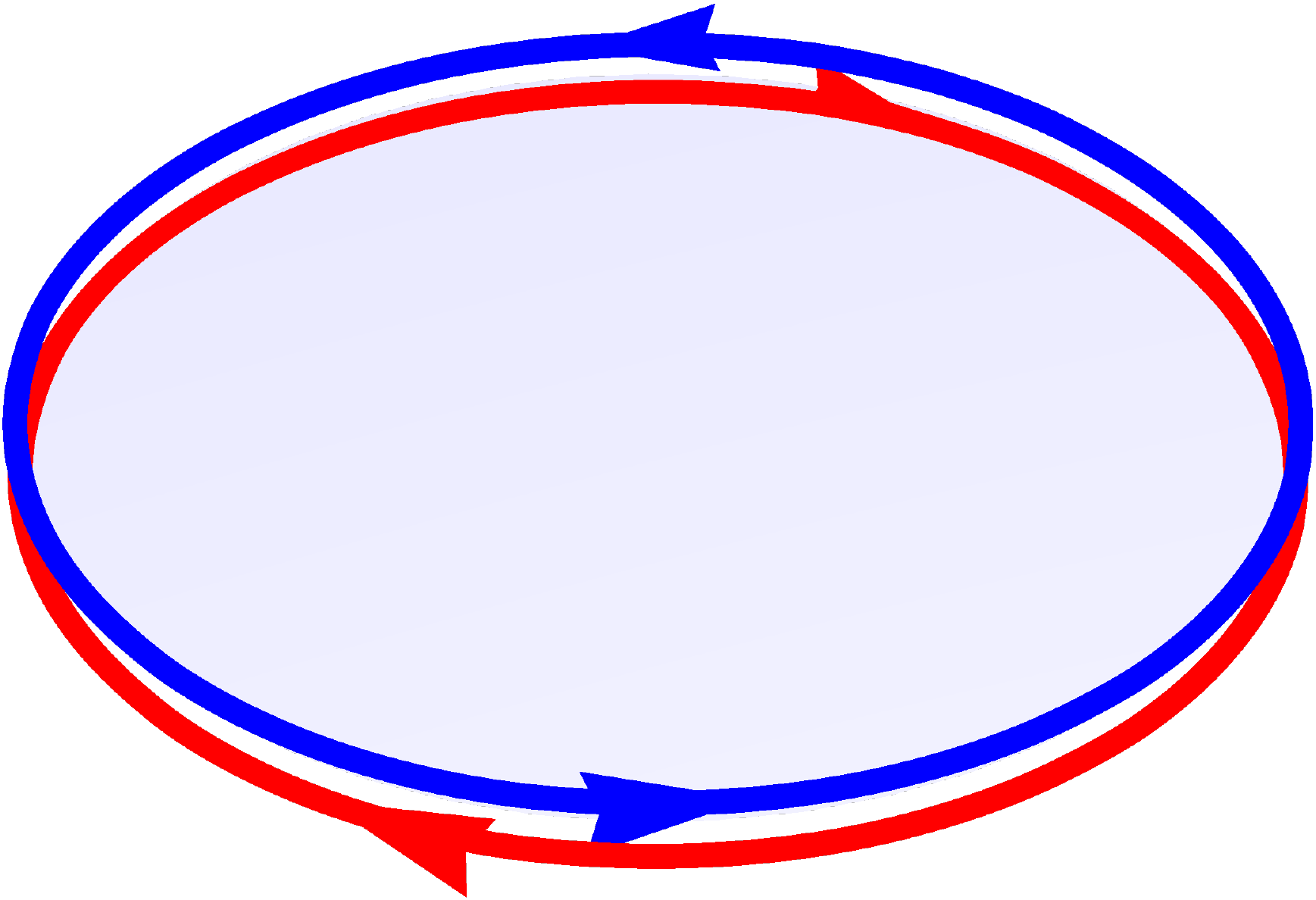}&\includegraphics[width=1cm, valign=c]{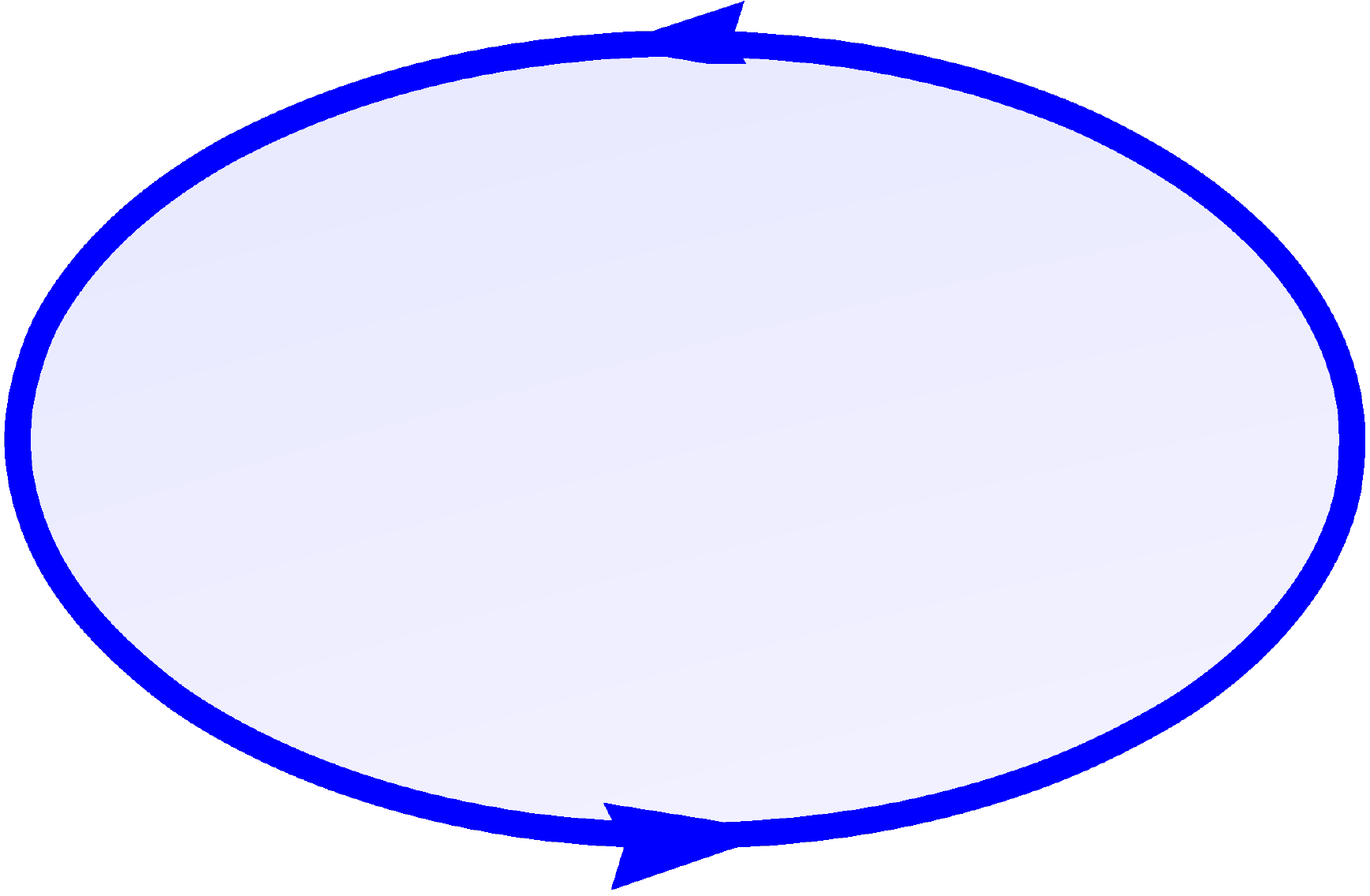} &  \includegraphics[width=1cm, valign=c]{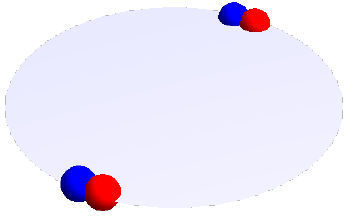} &  \includegraphics[width=1cm, valign=c]{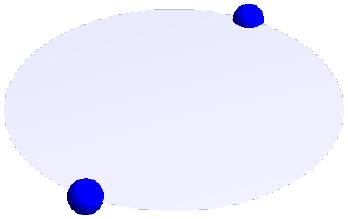} & & \\
			\hline
			1 &  $\mathbb{Z}_2$ & \includegraphics[valign=c]{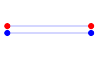}  &\includegraphics[valign=c]{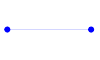} & &  & &
		\end{tabular}
	\caption{Illustration of the \emph{minimal} topology indicated by $\nu_{d,n,\xi} = 1$ with surface ($n\!=\!1$), hinge ($n\!=\!2$), and corner Majorana modes ($n\!=\!d$). In class DIII in 3D and class D in 2D, which have $\mathbb{Z}$ topological invariants, the indicators can not preclude first-order phases. Note that class D in 3D does not allow for a topological invariant in the ten-fold way, but $\nu_{3,1,D} = 1$ indicates Weyl modes in the bulk in a translationally symmetric system, .}	
	\label{tab:main}
 \end{table}

\textit{Main result} --
Our main finding is that for odd-parity order parameters, the symmetry indicator
\begin{equation}\label{eq:indicator}
	\kappa_{d,\xi}=r_{\xi} \sum_{\bs{k}\in \rm{TRIM}} \left( n^+_{\bs{k},\rm{N}}-n^-_{\bs{k},\rm{N}}\right) \in \mathbb{Z}
\end{equation}
is a $\mathbb{Z}_{2^d}$ valued strong topological invariant for $d$-dimensional weak-coupling superconductors protected by inversion symmetry in symmetry class $\xi\in\{\mathrm{D},\mathrm{DIII}\}$.
Here, $n^{\pm}_{\bs{k},\rm{N}}$ is the number of occupied states with inversion eigenvalue $\pm1$ at the time-reversal-symmetric momentum (TRIM) $\bs{k}$ in the normal-state band structure~\footnote{These are the momenta whose little group contains inversion, namely $k\in\{0, \pi\}$ in 1D, $\kk \in \{(0,0)$, $(0, \pi)$, $(\pi, 0)$, $(\pi, \pi)\}$ in 2D, and $\kk \in \{(0,0,0)$, $(0,0,\pi)$, $(0, \pi, 0)$, $(\pi, 0, 0)$, $(0, \pi, \pi)$, $(\pi, 0, \pi)$, $(\pi, \pi, 0)$, $(\pi, \pi, \pi)\}$ in 3D.} and the coefficients $r_{\mathrm{D}}=1$, $r_{\mathrm{DIII}}=1/2$. Note that in 3D for a normal state with TRS (class AII), $\kappa_{3,\mathrm{DIII}} = 2\kappa_1$, with $\kappa_1$ defined as in Ref.~\onlinecite{khalaf:2018b}.
Further, this indicator can be decomposed as
\begin{equation}
	\kappa_{d,\xi} \; \mathrm{mod}\ 2^d =\sum_{n=1}^d\nu_{d,n,\xi}\,2^{n-1},
	\label{eq:decomposition}
\end{equation}
where $\nu_{d,n,\xi}\in\{0,1\}$ are $\mathbb{Z}_2$ indices for $n$-th order topology in $d$ dimensions. 
For $\nu_{d,n,\xi}=1$, the `minimal' topology of the system is an $n$-th order topological phase, as illustrated in Tab.~\ref{tab:main}~\footnote{Note the similarity to the subgroup structure introduced in Ref.~\cite{trifunovic:2019}}.
The indices $\nu_{d,1,\xi}=1$ pertain to topological phases that do not require inversion for their protection.
The indices $\nu_{2,2,\xi}$ and $\nu_{3,3,\xi}$ describe higher order TSCs with corner Majorana modes and Kramer's pairs thereof. Finally, $\nu_{3,2,\xi}\!=\!1$ indicates a second-order TSC with one-dimensional chiral (helical) Majorana edge modes in class D (DIII).  
Note that cases of $d\!=\!n$ are not captured by an indicator based on a pure BdG formulation~\cite{ono:2018tmp}.

\textit{Atomic limit and symmetry indicators} --
While a quadratic Hamiltonian describing a non-interacting insulator is written in terms of electronic wave functions and thus possesses a natural atomic limit, no such limit exists for BdG Hamiltonians. In the latter, Bogoliubov quasi-particles describe electron-hole superpositions. Our strategy is thus to find a BdG Hamiltonian, where $\Delta_{\kk} \rightarrow 0$ does not close a gap and whose normal state possesses an AL in the sense of Ref.~\onlinecite{khalaf:2018b}. We then ask whether we can connect this Hamiltonian, which we denote as the AL of the superconductor, adiabatically to the original BdG Hamiltonian, without closing a gap. 

Since for a quadratic Hamiltonian (with the chemical potential $\mu$ equal to $\epsilon_{\rm F}$, the Fermi energy), the normal-state band structure has to be bounded from below, we can move $\mu$ below the lowest band to find such an AL Hamiltonian (we write in the following $\mu\!=\!-\infty$). 
To quantify the mismatch between the two Hamiltonians, we introduce a symmetry indicator as the difference in the number of irreducible representations of the occupied BdG bands $n^\alpha_{\bs{k},\rm{BdG}}|_{\mu=\epsilon_{\rm F}}$ (analog to $\kappa_1$~\cite{khalaf:2018b}) and of the occupied bands corresponding to the AL $n^\alpha_{\bs{k},\rm{BdG}}|_{\mu=-\infty}$. Specifically, the (inversion-) symmetry indicator is defined as
\begin{equation}\label{eq:bdg_kappa}
	\kappa_{d,\xi} = r_\xi \sum_{\bs{k}\in \rm{TRIMs}}\sum_{\alpha \in \{\pm\}}\alpha( n^\alpha_{\bs{k},\rm{BdG}}|_{\mu=\epsilon_{\rm F}}-n^\alpha_{\bs{k},\rm{BdG}}|_{\mu=-\infty})\,,
\end{equation}
where $\alpha $ refers to the inversion eigenvalue. The prefactors $r_{\mathrm{D}}=1$ and $r_{\mathrm{DIII}}=1/2$ guarantee that $\kappa_{d,\xi}\in\mathbb{Z}$.

To calculate this difference, we first need to understand the irreducible BdG band representations.
Given the normal-state Hamiltonian $H_{\kk}$ and the (mean-field) superconducting order parameter $\Delta_{\kk}$, the superconducting state can be described using the BdG Hamiltonian
\begin{equation}\label{bdg_hamiltonian}
	H^{\rm{BdG}}_{\bs{k}}=\Big(
		\begin{matrix}  
		H_{\bs{k}} & \Delta_{\bs{k}} \\
		\Delta^\dag_{\bs{k}}&-H_{-\bs{k}}^* \\
		\end{matrix}\Big)\,.
\end{equation}
Under any element of the generating point group, $g\in \mathcal{G}$, this Hamiltonian transforms as a scalar with transformations given by
$U^{\rm{BdG}}_{\bs{k}}(g)H^{\rm{BdG}}_{\bs{k}}U^{\rm{BdG}}_{\bs{k}}(g)^\dagger=H_{g{\bs{k}}}^{\rm{BdG}}$,
where
\begin{equation}\label{eq:symmetry_action}
	U^{\rm{BdG}}_{\bs{k}}(g)=\Big(
		\begin{matrix}
		U_{\bs{k}}(g) & 0\\
		0&\chi_g U^*_{-\bs{k}}(g) \\
	\end{matrix}\Big)\,.
\end{equation}
Note that $\chi_g\in U(1)$ is the eigenvalue of the order parameter under $g$, 
$
U_{\bs{k}}(g)\Delta_{\bs{k}}U^T_{\bs{-k}}(g)=\chi_g\Delta_{g\bs{k}},
$
in other words the character of the irreducible representation to which $\Delta_{\kk}$ belongs~\footnote{For higher-dimensional irreducible representations, a gap function is chosen that breaks additional (normal-state) symmetries. In the thus reduced symmetry group, the gap function belongs to a one-dimensional irreducible representation.}.
Finally, $H^{\text{BdG}}_{\bs{k}}$ obeys particle-hole symmetry $\mathcal{P}$ by construction, with $\mathcal{P}= \tau_x K$ acting in Nambu space and $\tau_i$ the Pauli matrices. 

In this work, we focus on systems with inversion symmetry $\mathcal{I}$.
At TRIMs, each eigenstate of $H_{\kk}^{\rm BdG}$ then either belongs to an even or odd irreducible representation, corresponding to an inversion eigenvalue $\alpha_{\kk}=\pm1$. Further, the order parameter is also either even or odd under inversion, meaning $\chi_{\mathcal{I}} = \pm 1$.

For the aforementioned construction, we then need $n^\alpha_{\kk,\rm BdG}$, the number of the occupied BdG bands transforming as the irreducible representation $\alpha$ at TRIM $\bs{k}$.
These transformation matrices can be deduced from the normal state transformation properties by assuming we are in the weak-coupling limit. In this limit, the transformation properties of the BdG bands at the TRIMs do not change for $\Delta_{\kk} \rightarrow 0$. Then, the transformation behavior of the BdG bands is entirely specified by the normal state properties and the irreducible representation of the order parameter. It follows from Eq.~\eqref{eq:symmetry_action} that  the irreducible representations 
of eigenstates of $H^*_{\kk}$ are given by 
$\chi_{\mathcal{I}} \alpha_{-\kk}$ \cite{ono:2018tmp}. 
Each eigenstate of the normal state Hamiltonian at a TRIM with energy $E_{
\kk}$ and eigenvalue $\alpha_{\kk}$ is thus mapped to two eigenstates in the BdG Hamiltonian, namely 
\begin{equation}
	(E_{\kk}, \alpha_{\kk}) \leftrightarrow \left \{ \begin{array}{l}(E_{\kk}, \alpha_{\kk})\\(-E_{\kk}, \chi_{\mathcal{I}}\alpha_{\kk})\end{array}\right. .
	\label{eq:mapping}
\end{equation}

For the difference in the number of irreducible representations in the BdG bands between $\mu=\epsilon_{\rm F}$ and $\mu = -\infty$ as defined in Eq.~\eqref{eq:bdg_kappa}, only occupied states of the original normal state are relevant.
In particular, Eq.~\eqref{eq:bdg_kappa} quantifies the difference of the occupied bands stemming from the normal (electron) states and the down-folded (hole) states.
It follows then from Eq.~\eqref{eq:mapping} that $\kappa_{d, \xi} \equiv 0$ for an inversion-even order parameter, $\chi_{\mathcal{I}} =1$, because $\kappa_{d,\xi}$ becomes independent of the chemical potential. For inversion-odd order parameters, on the other hand, we find Eq.~\eqref{eq:indicator}.
 
Let us inspect the stability of $\kappa_{d,\xi}$ against adding trivial AL bands below the Fermi level. Such additions should not change the topology of the system. Specifically, in 1D adding a fully filled trivial band below the chemical potential, in other words a band with $r_\xi \sum_{k} (n^+_{k}-n^-_{k})=2n$ ($n \in \mathbb{Z}$), will change the index by $2n$, making $\kappa_{1,\xi}$ a $\mathbb{Z}_2$ quantity. In general, $r_\xi \sum_{\kk} (n^+_{\kk}-n^-_{\kk})=2^dn$ in $d$ dimensions from which it follows that $\kappa_{d,\xi}$ is a $\mathbb{Z}_{2^d}$ quantity.

\begin{figure}[tb]
	\includegraphics[width=8cm]{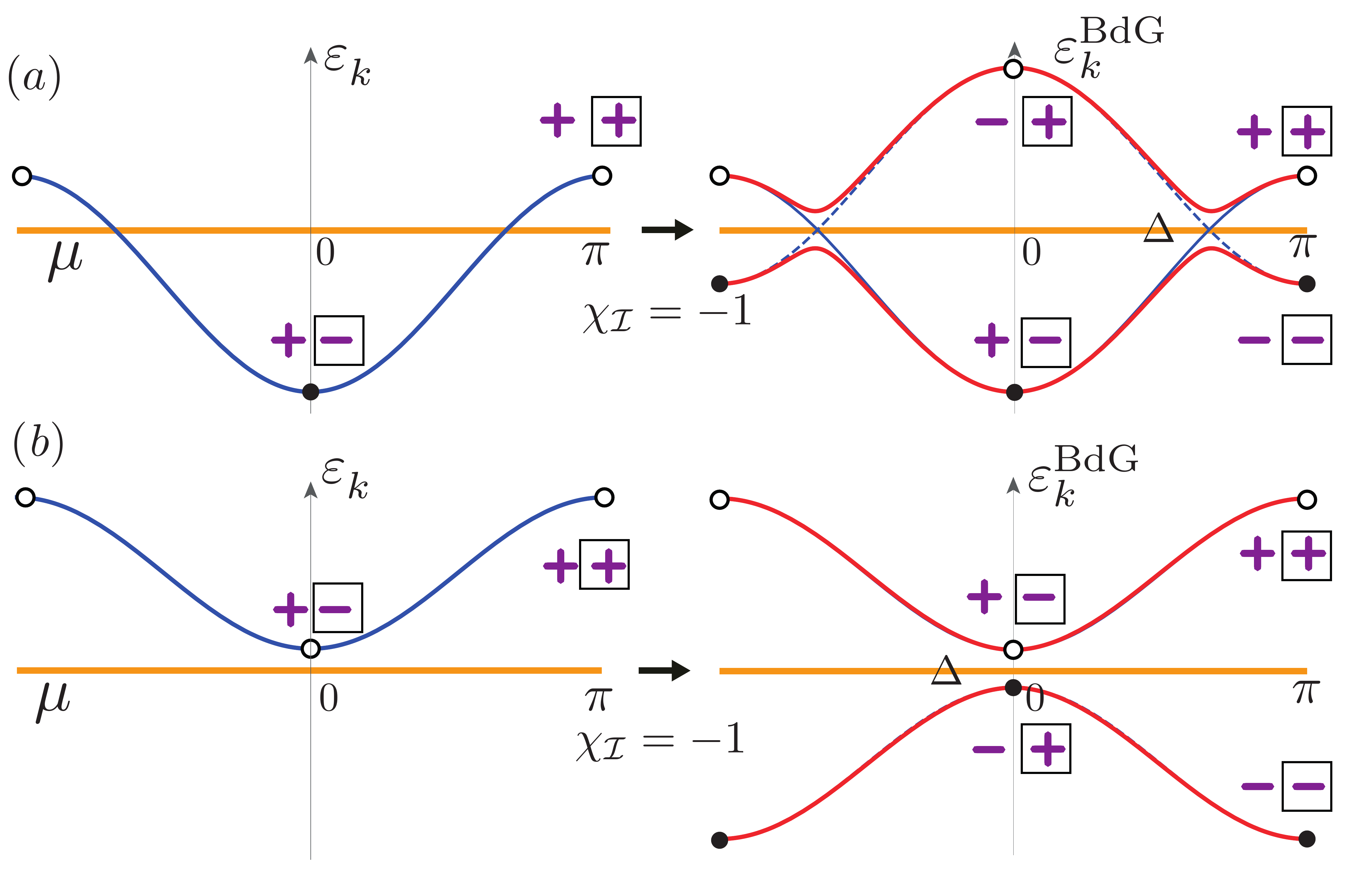}
	\caption{The BdG spectrum $\varepsilon^{\text{BdG}}_{k}$ obtained from a partially filled (empty) normal band $\varepsilon_{k}$ and an inversion-odd order parameter $\Delta_k$. The inversion eigenvalues at TRIMs $k=\{ 0,\pi\}$ for two different inversion representations $\mathcal{I}=\tau_z$ and $\mathcal{I}=-e^{ik}\tau_z$ are indicated by unboxed and boxed symbols $\pm$, respectively.}
	\label{fig:oneD_comparison}
\end{figure}  

\textit{One-dimensional case} --
To illustrate the importance of an AL for symmetry indicators of BdG bands, we consider the example of a 1D chain of spinless fermions.
The normal-state Hamiltonian in momentum space is given by a tight-binding nearest-neighbor model with dispersion $\varepsilon_k=-2t\cos(k)-\mu$ with hopping parameter $t$.
For spinless fermions, the order parameter is necessarily odd, meaning $\Delta_{k}=-\Delta_{-k}$ or ${\chi_{\mathcal{I}} \!=\! -1}$. Then, the BdG Hamiltonian reads 
\begin{equation}\label{eq:one_dimentional_bdg}
	H^{\rm{BdG}}_{k}=\varepsilon_k\tau_z+\Delta_{k}\tau_x\,.
\end{equation}  
Hamiltonian~\eqref{eq:one_dimentional_bdg} has inversion $\mathcal{I}$ with $\mathcal{I}H^{\rm{BdG}}_k \mathcal{I}^{-1}=H^{\rm{BdG}}_{-k}$
and belongs to class D. In 1D, class D is characterized by a $\mathbb{Z}_2$ topological invariant. 
Note that this invariant does not depend on the symmetry properties of the underlying normal-state bands or the presence and nature of fully occupied bands. In particular, the system is topologically non-trivial if an odd number of Fermi surfaces exists~\cite{sato:2010}.

Figure~\ref{fig:oneD_comparison} illustrates the ambiguity of the symmetry eigenvalues of the normal and BdG band structures for the topologically trivial and non-trivial cases.  
There are four possibilities for the representation of inversion symmetry of a single nondegenerate band depending on the symmetry and location of the Wannier orbitals compared to the inversion center.
In particular, when their locations coincide, the representation of inversion is given by $\mathcal{I}=\pm\tau_z$, with $\pm$ stemming from the symmetry of the underlying orbitals, for example $s$ and $p$ orbitals, respectively, and $\tau_z$ follows from Eq.~\eqref{eq:symmetry_action}. If the location of the orbitals and the inversion center are shifted by half a lattice constant, the inversion transformation is given by $\mathcal{I}=\pm\tau_z e^{ik}$ which is non-trivial in momentum space. 
For clarity, we only show the normal-state case of $\alpha_k = 1$ and $\alpha_k = - e^{ik}$ in Fig.~\ref{fig:oneD_comparison}. Then, the resulting BdG eigenvalues are exactly opposite for the two cases. Clearly, focusing solely on the BdG eigenvalues cannot indicate a topological phase in 1D. Calculating the indicator as given in Eq.~\eqref{eq:indicator}, however, distinguishes the topological phase. For completeness, Tab.~\ref{tab:oneD_indicators} summarizes all possible BdG eigenvalues and their connection to the (non-) trivial topology.

While we have only discussed the case without TRS above, adding TRS in the presence of inversion  simply doubles the number of bands. 
Since both cases in 1D have a $\mathbb{Z}_2$ topological invariant, $\kappa_{1,\xi}$ fully captures the topological nature of the phase for inversion symmetric systems.

\begin{table}[t]
	\centering
		\begin{tabular}{c!{\vrule width 1.25pt}M{0.6cm}|M{0.6cm}!{\vrule width 1.25pt}M{0.6cm}|M{0.6cm}!{\vrule width 1.25pt}M{0.6cm}|M{0.6cm}!{\vrule width 1.25pt}M{0.6cm}|M{0.6cm}} 
		         & \multicolumn{8}{c}{$\alpha_{k}$}\\
			 &\multicolumn{2}{c!{\vrule width 1.25pt}}{$\mu<-2|t|$}  & \multicolumn{4}{c!{\vrule width 1.25pt}}{$-2|t|<\mu<2|t|$}  & \multicolumn{2}{c}{$\mu>2|t|$} \\
			$\mathcal{I}$  & $0$ & $\pi$ & $0$ & $\pi$ & $0$ & $\pi$ & $0$ & $\pi$ \\
			\noalign{\hrule height 1.25pt}
			$\tau_z$  & $-$ & $-$ & \cellcolor{lightgray}$+$ & $-$ & $-$ & \cellcolor{lightgray}$+$ & \cellcolor{lightgray}$+$ & \cellcolor{lightgray}$+$ \\
			\hline
			$-\tau_z$  & $+$ & $+$ & \cellcolor{lightgray}$-$ & $+$ & $+$ & \cellcolor{lightgray}$-$ & \cellcolor{lightgray}$-$ & \cellcolor{lightgray}$-$ \\
			\hline
			$\tau_z e^{ik}$  & $-$ & $+$ & \cellcolor{lightgray}$+$ & $+$ & $-$ & \cellcolor{lightgray}$-$ &\cellcolor{lightgray} $+$ & \cellcolor{lightgray}$-$ \\
			\hline
			$-\tau_z e^{ik}$  & $+$ & $-$ & \cellcolor{lightgray}$-$ & $-$ & $+$ & \cellcolor{lightgray}$+$ & \cellcolor{lightgray}$-$ & \cellcolor{lightgray}$+$ \\
			\noalign{\hrule height 1.25pt}
			$\nu_{1,1,\mathrm{D}}$  & \multicolumn{2}{c!{\vrule width 1.25pt}}{$0$} & \multicolumn{2}{c!{\vrule width 1.25pt}}{$1$} & \multicolumn{2}{c!{\vrule width 1.25pt}}{$1$} & \multicolumn{2}{c}{$0$}  
		\end{tabular}
		\caption{Possible inversion eigenvalues $\alpha_k$ at TRIMs $k=\{0,\pi\}$ for the single-band BdG Hamiltonian Eq.~\eqref{eq:one_dimentional_bdg} (class D). Shaded cells indicate occupied normal states. The resulting $\nu_{1,1,\mathrm{D}}=\kappa_{1,\mathrm{D}}\,\mathrm{mod}\, 2$ correctly identifies the topological phase. Note that $\mu<-2t$ corresponds to the atomic limit defined in the text.}	
		\label{tab:oneD_indicators}
\end{table}
 
 \textit{Two-dimensional case.} -- In 2D without TRS, class D, the strong topological phase hosts chiral Majorana edge states and according to the AZ classification is characterized by a Chern number $C\in\mathbb{Z}$. Thus, the indicator only identifies directly first-order topological phases with odd $C$,  $C\mathrm{mod}2 = \nu_{2,1,\mathrm{D}}$.
  In class DIII, helical Majorana edge states exist in the topological phase and the $\mathbb{Z}_2$ invariant characterizing this first-order topological phase~\cite{ryu:2010} coincides with the indicator $\nu_{2,1,\mathrm{DIII}}$ for systems with inversion symmetry.

When $\nu_{2,1,\mathrm{D}}=0$, the system may still be in a strong topological phase with even but nonzero Chern number. Assuming $C=0$ instead, $\nu_{2,2,\mathrm{D}}=1$ indicates a second order topological superconductor with two corner Majorana states in an inversion-symmetric geometry. 
This higher-order phase is thus the `minimal' topology required by the index $\nu_{2,2,\mathrm{D}}=1$.
One can obtain such a superconductor from a strong topological superconductor with $\nu_{2,1,\mathrm{DIII}}=1$  by introducing a TRS-breaking mass term, while preserving inversion symmetry \cite{khalaf:2018a}.  

A TRS second-order TSC is indicated by $\nu_{2,2,\mathrm{DIII}}=1$ and features Kramers pairs of Majorana corner modes. It can be obtained by adding the fully occupied band of a $\mathbb{Z}_2$ topological insulator in symmetry class AII to an otherwise trivial odd-parity superconductor.

\textit{Three-dimensional case} --
The strong TSC in 3D with TRS and PHS belong to class DIII and are characterized by the winding number $\text{W}\in\mathbb{Z}$~\cite{ryu:2010}. As with the Chern number in 2D, the indicator only identifies directly first-order topological phases with odd $W$,  $W\,\mathrm{mod}\,2 = \nu_{3,1,\mathrm{DIII}}$.
In the AZ classification, class D has no gapped topological phase, but given translational symmetry, $\nu_{3,1,\mathrm{D}}\!=\!1$ indicates momentum-space planes with different Chern numbers and thus a gapless Weyl superconductor \cite{meng:2012,fischer:2014}.

When  $\nu_{3,1,\mathrm{DIII}}=0$, the system may still be in a strong topological phase with even but nonzero winding number $W$. Assuming $W=0$ instead, $\nu_{3,2,\mathrm{DIII}}=1$ indicates a second order topological superconductor with a Kramer's pair of helical hinge Majoranas in an inversion-symmetric geometry. Further, $\nu_{3,2,\mathrm{DIII}}=0$ but $\nu_{3,3,\mathrm{DIII}}=1$
 indicates a third-order topological superconductor with a Kramer's pair of Majoranas corner modes.
These higher-order phases are again the minimal topology compatible with the respective values of the symmetry indicators. The second and third-order phases can be obtained by adding the fully occupied band of, respectively, a 3D $\mathbb{Z}_2$ topological insulator in symmetry class AII and a higher-order topological insulator with $\kappa_1\,\mathrm{mod}\,4=2$  to an otherwise trivial odd-parity superconductor.

We close by constructing examples of the bulk-boundary correspondence of all higher-order phases in class D and DIII using surface Dirac theories.
For this purpose, we follow Ref.~\cite{khalaf:2018a} and start from a (first-order) strong topological phase in DIII characterized by a non-zero winding number $W=\pm1$ and described by the bulk Hamiltonian $\mathcal{H}_{\pm}$ for an odd-parity superconductor with a single Fermi pocket enclosing the $\Gamma$ point in the normal state. This Hamiltonian is gapped in the bulk but has a Majorana surface cone governed by
\begin{equation}
	h_{\pm}=\mp(\bs{k}\times \bs{n}_{\bs{r}})\cdot \bs{\tau}\,,
	\label{eq:dirac}
\end{equation}
where $\bs{\tau}$ are Pauli matrices in particle-hole space and $\bs{n}_{\bs{r}}$ is the normal vector to the surface. Furthermore, the surface theory possess PHS, TRS and inversion, which on the surface are given by $\mathcal{T}=\tau_y\mathcal{K}$, $\mathcal{P}_{\bs{r}}=\mp\,\bs{n}_{\bs{r}}\cdot \bs{\tau}\tau_y\mathcal{K}$, and $\mathcal{I}=\mp\tau_0$. Given a fixed bulk normal-state inversion eigenvalue of the occupied band at $\Gamma$, the sign of the winding number determines the sign of the induced inversion symmetry on the surface~\cite{khalaf:2018a}. 

We can arrive at all possible higher-order phases by adding copies with opposite winding numbers (class DIII) and breaking TRS (class D).
A second-order TSC in class D is obtained by adding to Eq.~\eqref{eq:dirac} the only $\kk$-independent TRS-breaking term that preserves PHS, $\tilde{M}_{\bs{r}} \bs{n}\cdot \bs{\tau}$. Inversion symmetry further requires $\tilde{M}_{\bs{r}}=-\tilde{M}_{-\bs{r}}$, which implies two domains with opposite mass sign of this surface Dirac theory. Along the domain wall, which is an inversion symmetric path on the surface, a chiral mode remains~\footnote{\label{f:2D}Note that the same Hamiltonian at $k_z=0$ describes a 2D second-order TSC in class D}.

The class DIII second-order phase is built by taking two copies with the opposite winding numbers $W$, $\mathcal{H} = \mathcal{H}_{+}\oplus\mathcal{H}_{-}$, which has $W=0$. The surface theory of such a system can be written as 
\begin{equation}
	h_2=-(\bs{k}\times \bs{n}_{\bs{r}})\cdot \bs{\tau} \gamma_z\,,
	\label{eq: h2}
\end{equation}
where the Pauli matrices $\bs{\gamma}$ act on the $+/-$ grading.
We can add a mass term $M_{\bs{r}}\gamma_x$, which preserves all the symmetries of the Hamiltonian, namely PHS, TRS and inversion, iff $M_{\bs{r}}=-M_{-\bs{r}}$. For the surface theory, these symmetries are represented by $\mathcal{T}=\tau_y\mathcal{K}$, $\mathcal{P}_{\bs{r}}=-\gamma_z\,\bs{n}_{\bs{r}}\cdot \bs{\tau}\tau_y\mathcal{K}$ and $\mathcal{I}=-\gamma_z$, respectively. This form of $\mathcal{I}$ follows from the requirements $W=0$ and $\nu_{3,2,\mathrm{DIII}}=1$.
The fact that $M_{\bs{r}}$ is an odd function on the surface enforces again two domains with opposite mass sign. A gapless Kramer's pair of helical Majorana modes is bound to the domain wall, which is an inversion symmetric path on the surface.

The third order TSC in D class can be obtained from this second order TSC in class DIII by adding a TRS breaking mass term $\tilde{M}_{\bs{r}}\gamma_y$ with $\tilde{M}_{\bs{r}}=-\tilde{M}_{-\bs{r}}$ to the Hamiltonian, which would correspond to applying a magnetic field, for instance. This gaps the helical modes except for two inversion-related points on the surface, where two corner Majorana modes remain. 

Finally to build third order TSC with TRS we need to combine four copies of $\mathcal{H}_{\pm}$ which total to $W=0$, i.e., $\mathcal{H}_{+}\oplus\mathcal{H}_{-}\oplus\mathcal{H}_{+}\oplus\mathcal{H}_{-}$. The surface theory of this TSC is given by 
\begin{equation}
h_4=-(\bs{k}\times \bs{n}_{\bs{r}})\cdot \bs{\tau} \gamma_z \sigma_0\,,
\end{equation} 
where we introduced  the Pauli matrices $\bs{\sigma}$ which act on the additional grading introduced as compared to Eq.~\eqref{eq: h2}. This system has PHS, TRS and inversion given be $\mathcal{T}=\tau_y\sigma_0\mathcal{K}$, $\mathcal{P}_{\bs{r}}=-\gamma_z\,\bs{n}_{\bs{r}}\cdot \bs{\tau}\tau_y\sigma_0\mathcal{K}$ and $\mathcal{I}=-\gamma_z\sigma_0$. This form of $\mathcal{I}$ follows from the requirements $W=0$, $\nu_{3,2,\mathrm{DIII}}=0$, and $\nu_{3,3,\mathrm{DIII}}=1$.
Since each of the four constituent strong TSC of the DIII class is characterized by a Majorana surface cone, we can combine these strong TSC in the pairs by adding a mass term of the form  $M_{\bs{r}}\gamma_x\tau_0\sigma_0$, with $M_{\bs{r}}=-M_{-\bs{r}}$, which gaps out the surface states, but leaves two inversion-symmetric gapless lines carrying a helical Majorana mode each. Then, another mass term given by $\tilde{M}_{\bs{r}}\gamma_y\sigma_y$ with $\tilde{M}_{\bs{r}}=-\tilde{M}_{-\bs{r}}$ can be added, which anticommutes with the Hamiltonian and preserves all its symmetries, to gap out the Majorana edge modes but leave the Majorana corner modes at inversion-symmetric points.

\textit{Conclusions} --
We presented a unified symmetry indicator to determine the topology of spin-orbit coupled TSCs with inversion symmetry and odd-parity order parameter. The indicator can be evaluated from the knowledge of the normal state band structure alone and describes in particular all higher order topological superconductors with hinge and corner Majorana modes. Our analysis is based on the definition of a physical atomic limit for superconductors and can be generalized to other spatial symmetries besides inversion.

\textit{Note Added} --- 
In completing this work we became aware of Ref.~\cite{ahn:2019tmp}, which primarily discusses symmetry indicators for TSCs in symmetry class BDI and then generalizes to symmetry class D. Our results agree where they overlap.

\textit{Acknowledgements} --- 
We thank Piet Brouwer, Luka Trifunovic, Ronny Thomale and Frank Schindler for stimulating discussions.  
This project has received funding from the European Research Council (ERC) under the European Union's Horizon 2020 research and innovation programm (ERC-StG-Neupert-757867-PARATOP).

\bibliography{biblio}
\clearpage
\end{document}